\documentclass[12pt]{article}
\usepackage{latexsym}

\def\cQ{{\cal Q}}
\def\cS{{\cal S} }

\def\cC{{\cal C}}
\def\ket#1{\mid~\!\!\!{#1}~\!\!\rangle}
\def\bra#1{\langle~\!\!{#1}~\!\!\!\mid}
\def\+-{\buildrel + \over -}
\def\IF {if and only if }

\def\qm{quantum mechanics}
\def\Q{quantum }
\def\q{quantum}

\def\cR{{\cal R}}

\def\${\enskip$}
\def\M{measurement }
\def\m{measurement}

\def\mi{measuring instrument }

\begin{document}

{\bf\large \noindent General Theory of Over\M \\ of Discrete Quantum
Observables\\ and Application to Simultaneous Measurement}

\vspace{1.3cm}

{\bf\noindent Fedor Herbut}

\vspace{1.3cm}

\rm

{\bf\noindent Abstract.} A complete theory of over\M by measuring refinements of observables is presented. It encompasses a wider set of functions of observables (coarsenings) . Thus the theory has a broad potential application.It is applied to a thorough investigation of simultaneous \m s. In particular, the set of all simultaneous \m s for a given pair of compatible observables is determined.\\

\noindent {\bf Keywords} Measurement. Functions of observables. Compatible observables.

\vspace{0.5cm}

\section{Introduction}

\noindent
It is a textbook claim that two compatible discrete observables, i. e., ones of which the Hermitian operators representing them commute and have no continuous parts in their spectra, can be simultaneously measured. And this is done, so it is further claimed, by finding a common eigenbasis of the two operators and by measuring in which  of the basis states the system is. The eigenvalues of the two operators that correspond to the measured basis state are then the  simultaneous results of the measurement. This is a superficial and incomplete but typical presentation of simultaneous \m .

The two discrete observables are {\it overmeasured} by a common over\m , though this term is usually not used. In this article a {\it complete theory of over\M } is expounded together with a complete theory of simultaneous \m s as an application.\\
{\footnotesize \rm \noindent
\rule[0mm]{4.62cm}{0.5mm}\\
\noindent F. Herbut\\
Serbian Academy of Sciences and Arts,
Knez Mihajlova 35, 11000 Belgrade,
Serbia\\
e-mail: fedorh@sanu.ac.rs\\
If one defines general exact \m , following \cite{BLM}, by the {\it calibration condition} (see relation (4) below), then over\M is the {\it most general exact \m }. The opposite of over\M is under\m . Since the points of a continuous spectrum cannot be measured, they must be undermeasured. Von Neumann in his famous book \cite{vonNeum} (cf chapter III, section 3. p. 220 there) explains this, though he does not use the term "under\m ". (His term for under\M is "\M with only limited accuracy".)

It is hoped that the complete theory of over\M that is to be presented will not only give a deeper conceptual insight in \M theory, but also find new applications.

The investigation is restricted to {\it discrete observables} in this article. They will always be given in their {\it unique spectral form} (unless otherwise stated), which means, by definition, that there is no repetition in the eigenvalues \$\{o_k:\forall k\}\$ that are displayed in the spectral form:
$$O_A=\sum_ko_kE_A^k, \eqno{(1a)}$$
so that \$\{E_A^k:\forall k\}\$ are the corresponding  eigen-projectors. The index A denotes the measured subsystem. The spectral form is
accompanied by the spectral (orthogonal projector) decomposition of the identity operator \$I_A\$ (also called the "completeness relation")
$$\sum_kE_A^k=I_A. \eqno{(1b)}$$

When \$O_A\$ is measured  in a suitable interaction with a \mi B, then an initial or ready-to-measure state \$\ket{\phi}_B^i\$ together with a so-called {\it pointer observable}
$$P_B=\sum_kp_kF_B^k \eqno{(2)}$$
are given. The eigen-projectors \$\{F_B^k:\forall k\}\$ are metaphorically called |pointer positions".
Also they satisfy the completeness relation \$\sum_kF_B^k=I_B\$. Notice the co-indexing in (2) and (1a) based on a one-to-one relation between the possible \M results \$\{o_k:\forall k\}\$ and all possible pointer positions.

The suitable \M interaction is assumed to be incorporated in a unitary operator \$U_{AB}\$, which
maps the composite initial state to the final state
\$\ket{\Phi}_{AB}^f\$
$$\ket{\Phi}_{AB}^f=U_{AB}\Big(\ket{\phi}_A^i
\ket{\phi}_B^i\Big),\eqno{(3)}$$
where \$\ket{\phi}_A^i\$ is an arbitrary initial state of the object subsystem.

This is the basic formalism of unitary \M theory, or pre\M theory or \M theory short of collapse
\cite{BLM}, \cite{FHArxiv}, \cite{FHSubsMeas}.
The general unitary (also called "exact") \m s of discrete observables are defined by the {\it calibration condition}, which requires that if the object has a sharp value \$o_{\bar k}\$ of the measured observanle in the initial state, then the final composite state has the corresponding sharp pointer position \$F_B^{\bar k}\$:
$$E_A^{\bar k}\ket{\phi}_A^i=\ket{\phi}_A^i
\quad\Rightarrow\quad F_B^{\bar k}
\ket{\Phi}_{AB}^f=\ket{\Phi}_{AB}^f.\eqno{(4)}$$
(Note that the mutually equivalent eigenvalue equations \$E_A^{\bar k}\ket{\phi}_A^i=\ket{\phi}_A^i\$ and \$O_A\ket{\phi}_A^i=o_{\bar k}\ket{\phi}_A^i\$ are the standard way to express certainty in \qm , and "$\Rightarrow$" stands for logical implication.)

In this study we will not treat the important special case of nondemolition (synonyms: repeatable, predictive, first-kind) \m s, nor the much used even more special special case of ideal \m s \cite{Lud}.\\

It is known from von Neumann's book \cite{vonNeum} that an observable \$O_A\$ given by (1a) can be measured by measuring a complete observable, i. e., one with no degeneracy in any of its eigenvalues,
$$O_A^r=\sum_k\sum_{n_k}o^r_{k,n_k}\ket{k,n_k}
\bra{k,n_k},\eqno{(5a)}$$
$$ (k,n_k)\not=(k',n'_{k'})\enskip\Rightarrow\enskip o^r_{k,n_k}\not= o^r_{k',n'_{k'}},\eqno{(5b)}$$
which is a so-called {\it refinement} of \$O_A\$, i. e., for which
$$\forall k:\quad \sum_{n_k}\ket{k,n_k}
\bra{k,n_k}=E_A^k \eqno{(5c)}$$ is valid.

One is dealing with over\M of \$O_A\$, where actually \$O_A^r\$ is measured, and, if. e. g., \$o^r_{k,n_k}\$ is the result of \m , then  by \Q -logical implication, due to \$\ket{k,n_k}\bra{k,n_k}\leq E_A^k\$ (symbolic for \$\ket{k,n_k}\bra{k,n_k}E_A^k =\ket{k,n_k}\bra{k,n_k}\$), also the pointer position \$E_A^k\$ has occurred or the result \$o_k\$ of \$O_A\enskip\Big(=\sum_{k'}o_{k'}E_A^{k'}\Big)\$ is obtained.\\

\section{General Theory of Over\M }

\noindent
The unitary \Q formalism is restricted to unitary evolutions, and, as well known, it cannot in general derive the (unknown) final state \$(\ket{\Phi}_{AB}^f)^k\$ of {\it complete \m }, which includes collapse to the definite result \$p_k\$ or, equivalently, the occurrence of the pointer position \$F_B^k\$. But the very fact that it
contains the information of a definite \$o_k\$ result,
i. e., due to \$(\bra{\Phi}_{AB}^f)^k F_B^k
(\ket{\Phi}_{AB}^f)^k=1\$, one must have equivalently,
$$(\ket{\Phi}_{AB}^f)^k=F_B^k (\ket{\Phi}_{AB}^f)^k.\eqno{(6)}$$ The final state (6) of complete \M might even be mixed. For simplicity we restrict it to a pure state.\\

\subsection{Over\M - The formal part}

{\it Over\M } is usually {\it defined in a more narrow sense} by any single-valued function \$f(\dots )\$ on the real axis. It determines an observable \$\bar O_A\$ that is the corresponding function of the given observable \$O_A\enskip\Big(=\sum_ko_k E_A^k\Big)$: $${\bar O}_A
\equiv f(O_A)\equiv \sum_kf(o_k){E_A^k}=$$ $$\sum_l\bar o_l{E_A^l},\qquad
l\not= l'\Rightarrow\bar o_l\not=\bar o_{l'}.\eqno{(7a)}$$ Note that the first spectral form in (7a), unlike the second one, is, in general, non-unique.

In the context of over\m , \${\bar O}_A\$ is called a coarsening or a coarser observable, and \$O_A\$ is said to be a refinement or a finer observable. (These terms are meant in the improper sense. For instance, "finer" is actually "properly finer" or equal.)

The indices \$l\$ are defined so as to make the spectral form of the coarser observable \${\bar O}\$ unique. This implies that the index set \$\{\forall k\}\$ in the unique spectral form of the finer observable \$O\$ is broken up into equivalence classes: \$\{\forall k\}=\sum_lC_l\$. In other words,  it can be viewed as the union of non-intersecting subsets (classes) \$C_l\$. Belonging to the same class \$C_l\$ is defined as follows.
$$\forall l:\enskip  k,k'\in C_l\quad\Leftrightarrow\quad
f(o_k)=f(o_{k'})=\bar o_l.\eqno{(7b)}$$
Thus \$f(\dots )\$, primarily given as a function on the real axis, determines a function, we denote it by the same symbol \$f\$, mapping the index set \$\{\forall k\}\$ onto the new index set \$\{\forall l\}\$. Note that the {\it inverse multivalued function} \$f^{-1}\$ takes the latter index set onto the former and its images are precisely the mentioned equivalence classes: $$\forall l: \enskip k,k'
\in C_l\quad\mbox{if and only if}\enskip  k,k'\in f^{-1}(l).\eqno{(7c)}$$

It is sometimes useful to define {\it overmeasurement in a broader sense} by an (arbitrary) single-valued map \$f\$ taking the index set \$\{\forall k\}\$ of the finer observable \$O\$ onto the index set \$\{\forall l\}\$ of the coarser observable \${\bar O}\$. But always the essential thing is the relation
$$\forall l:\quad E_A^l=\sum_{k,f(k)=l} E_A^k.\eqno{(8a)}$$ Relation (8a) follows from (7a) if one has the narrower definition, and it is the most important part of the definition of overmeasurement in the broader definition. In the latter case the eigenvalues \$\{\bar o_l\}\$ of the coarser observeble need not be related to those of the finer observable.

As a consequence of (8a), the orthogonality of the projectors \$\{E_A^k:\forall k\}\$ leads to
$$E_A^lE_A^k=0\quad\mbox{if}\quad
f(k)\not=l.\eqno{(8b)}$$\\

Parallelly with the unique spectral form of the coarser observable, also the unique spectral form
$${\bar P_B}=\sum_l\bar p_lF_B^l,\eqno{(9)}$$
of the pointer observable of the coarsening is going to play an important role. Note that the eigenvalues \$\{\bar p_l:\forall l\}\$ can be arbitrary distinct real numbers. Further, by definition
$$\forall l:\quad F_B^l=\sum_{k,f(k)=l} F_B^k,\eqno{(10a)}$$ where the function \$f:\enskip \{\forall k\}\rightarrow\{\forall l\}\$ is the one that determines the coarsening \$O\rightarrow{\bar O}\$. Relations (10a)  are symmetrical to (8a).

One has also
$$F_B^lF_B^k=F_B^k\quad\Big(
\Leftrightarrow\quad F_B^l\geq
F_B^k\Big)\quad\mbox{if}\quad f(k)=l\enskip .\eqno{(10b)}$$

Naturally, the eigen-projectors of the finer and of the coarser observable and the eigen-projectors of the corresponding pointer observables satisfy symmetrical relations. But we have written down only those that we shall make use of.\\

\subsection{Over\M - The physical part}

Now we make {\it the first physical step} showing that {\it any} unitary \M of an observable \$O_A\$ is by this very fact {\it a unitary \M also of any coarser observable \${\bar O}_A\$}
related to the finer observable \$O_A\$ by a given map of the index set of the latter onto that of the coarser observable. In particular, we shall demonstrate that the calibration condition, which is by definition valid for the \M of the finer observable, implies that also the calibration condition for the coarser observable is satisfied.

We assume that the initial state \$\ket{\phi}_A^i\$ of the object has a sharp value \$\bar o_{\bar l}\$
of the coarser observable:
 $$\ket{\phi}_A^i=E_A^{\bar l}\ket{\phi}_A^i.
\eqno{(11)}$$  Utilizing the completeness relation \$I_A=\sum_kE_A^k\$ in the  decomposition \$\ket{\phi}_A^i=\sum_kE_A^k\ket{\phi}_A^i\$ and (11), \$\ket{\Phi}_{AB}^f\$, which is defined by (3),  becomes equal to
$$\sum_k||E_A^k\ket{\phi}_A^i||
U_{AB}\Big[\Big(E_A^kE_A^{\bar l}\ket{\phi}_A^i\Big/ ||E_A^k\ket{\phi}_A^i||\Big)\ket{\phi}_B^i\Big].
\eqno{(12)}$$
Since the sum can be broken up \$\sum_k\dots =\sum_{k,f(k)\not=\bar l}\dots +\sum_{k,f(k)=\bar l}\$, (8b) makes the first sum zero. Hence, making use of the assumption that the \M of the finer observable satisfies the calibration condition in the form of inserting \$F_B^k\$, and using (11) again to suppress \$E_A^{\bar l}\$, \$\ket{\Phi}_{AB}^f\$ is further equal to
$$\sum_{k,f(k)=\bar l}
||E_A^k\ket{\phi}_A^i||F_B^k
U_{AB}\Big[\Big(E_A^k\ket{\phi}_A^i\Big/ ||E_A^k\ket{\phi}_A^i||\Big)\ket{\phi}_B^i\Big].
\eqno{(13)}$$

Finally, taking into account (10b), we obtain
$$F_B^{\bar l}\ket{\Phi}_{AB}^f= \ket{\Phi}_{AB}^f,\eqno{(14)}$$, which expresses certainty. This proves the claim. Thus, in view of (11) and (14), the calibration condition is valid for the over\M of the coarser observable.

Naturally, due to the usual convention, if \$E_A^k\ket{\phi}_A^i=0\$ in some term, then the expression that follows in the same term need not be defined; the term is by definition zero.\\

Now we can make {\it the second physical step} concerning {\it the result of complete \m }. The claim is that {\it if the complete \M of the finer observable produces the result \$o_k\$, then this same process of \M gives the result \$\bar o_{f(k)}\$ for the coarser observable.} The proof is an immediate consequence of (1ob). Namely, putting \$l\equiv f(k)\$, one obtains
$$F_B^l\{\ket{\Phi}_{AB}^f\}^k=
F_B^l\Big(F_A^k\{\ket{\Phi}_{AB}^f\}^k\Big)=
\Big(F_B^lF_A^k\Big)\{\ket{\Phi}_{AB}^f\}^k=$$  $$
F_A^k\{\ket{\Phi}_{AB}^f\}^k=
\{\ket{\Phi}_{AB}^f\}^k.\eqno{(15)}$$

We have thus proved that the final state
\$\{\ket{\Phi}_{AB}^f\}^k\$ of complete \M has the definite result \$\bar o_{l\equiv f(k)}\$ of the coarser observable. If this final state is mixed, the proof is analogous, but it requires certain generalizations of the formalism. Hence it is omitted for simplicity.\\

If a coarser observable \${\bar O}_A\$ in the proper sense is given first, there exist various refinements; there can even be refinements of refinements. And one has transitivity: a refinement of a refinement is a refinement of the coarsest observable \${\bar O}_A\$. Therefore one can speak of {\it degrees of over\M } of the given observable \${\bar O}_A\$.

The two {\it extreme degrees} are: {\it minimal \m }, when there is actually no refinement, and {\it maximal over\m }, when the measured finer observable \$O_A\$ is a complete observable, i. e., one all  eigenvalues of which are non-degenerate (cf the end of the Introduction).

The best known example of minimal \M is ideal \m , also called L\" uders or von Neumann-L\" uders \M (cf section 7 in \cite{FHArxiv}).

Minimal \M in a general sense was introduced by the present author \cite{FHMinMeas}. Maximal over\M is also called \M in a given basis (having in mind the eigen-basis of the complete observable; its eigenvalues anyway play no role in \M theory).

One should note that, if minimal \M is included in over\M (as the trivial, improper  extreme), then  every \M is an over\m .\\

\section{Simultaneous \M }

\noindent
This section is devoted to an illustration of application of over\M to a topic that is well known but not well proved and not well understood in its fine details.

To begin with, let us {\it define} that by {\it simultaneous \M } of two observables \$O_A'\enskip\Big(=\sum_mo_mE_A^m\Big)\$ and \$O_A''\enskip\Big(=\sum_no_nE_A^n\Big)\$ is understood \M of one observable \$O_A\enskip\Big(=\sum_ko_kE_A^k\Big)\$ that is so chosen that any result \$o_k\$ implies (by \Q -logical implication) a result \$o_{m(k)}\$ of \$O_A'\$ and simultaneously a result \$o_{n(k)}\$ of \$O_A''\$. Besides, each possible result \$o_m\$ of \$O_A'\$ and \$o_n\$ of \$O_A''\$ must be thus obtainable for some initial state \$\ket{\phi}_A^i\
$.\\

\subsection{Common over\M and
compatibility}

The very definition of simultaneous \M implies that , by {\it necessity}, there must exist two functions \$f'\$ and \$f''\$ mapping the set of all indices \$\{\forall k\}\$ onto the sets of all indices \$\{\forall m\}\$ and \$\{\forall n\}\$ respectively so that using the notation
$$\forall k:\quad f'(k)=m(k),\qquad
f''(k)=n(k),\eqno{(16a)}$$
one has
$$\forall k:\quad E_A^k\leq E_A^{m(k)}\quad
\Big(E_A^kE_A^{m(k)}=E_A^k\Big)\quad\mbox{and}$$  $$
E_A^k\leq E_A^{n(k)}\quad
\Big(E_A^kE_A^{n(k)}=E_A^k\Big).\eqno{(16b)}$$

It is seen that \$O_A\$ must be a {\it common refinement} of \$O_A'\$ and \$O_A''\$ and hence the
\M of \$O_A\$ a common over\M of the latter two observable. Thus, {\it necessity} of the common refinement claim is proved.\\

As to proving {\it sufficiency} of the stated claim, it clearly follows from the definition of simultaneous \M that any common over\M will achieve it.\hfill $\Box$\\

Furthermore, relations (16a,b) imply
$$\forall m:\quad E_A^m=\sum_{k\in (f')^{-1}(m) }E_A^k$$
$$\mbox{and}\quad \forall n:\quad E_A^n=\sum_{k\in (f'')^{-1}(n)}E_A^k,\eqno{(17)}$$
which, in turn, has
$$\forall m,n:\quad [E_A^m,E_A^n]=0\eqno{(18)}$$
as its consequence.

Two observables that satisfy the commutativity condition (18) are said to be {\it compatible}.
In this way it is proved that for simultaneous measurability compatibility is {\it necessary}.
We now prove that it is also {\it sufficient}.

Assuming the validity of (18), each product \$E_A^mE_A^n\$ is a projector, and $$(E_A^mE_A^n)
(E_A^{m'}E_A^{n'}=(E_A^mE_A^{m'})(E_A^n)E_A^{n'})=$$ $$
\delta_{m,m'}\delta_{n,n'}E_A^mE_A^n,$$ i. e., any two projectors in the set \$\{E_A^mE_A^n:\forall m,\forall n\}\$ are orthogonal. Finally, multiplying the two completeness relations \$\sum_mE_A^m=O_A\$ and \$\sum_nE_A^n=I_n\$, one obtains the ompleteness kation \$\sum_m\sum_nE_A^mE_A^n=I_A\$.

Let us enumerate by \$k\$ all non-zero distinct projectors
$$E_A^k\equiv E_A^mE_A^n\not= 0,\eqno{(19a)}$$ and take an arbitrary set \$\{o_k:\forall k\}\$ of distinct real numbers. Then, it is obvious from the arguments above, that
$$O_A\equiv\sum_ko_kE_A^k\eqno{(19b})$$
is a common refinement of \$O_A'\$ and
\$O_A''\$. Hence its \M is a common over\M of these two given observables. \hfill $\Box$\\

If two observables \$O_A'\$ and \$O_A''\$ are bounded, then they are compatible \IF they commute \$[O_A',O_A'']=0\$. Also this claim is, unlike its proof, well known. (For the reader's convenience we prove it in  Appendix B.)

Incidentally, it is known in linear analysis, or rather from the theory of at most countably infinite complex Hilbert spaces \cite{vonNeum}, that if the spectrum \$\{o_m:\forall m\}\$ of a given observable \$O_A'\enskip\Big(=\sum_mo_mE_A^m\Big)\$ is known, then a useful necessary and sufficient condition for boundedness of \$O_A'\$ is that the spectrum belongs to a finite closed interval:
$$\{o_m:\forall m\}\subset [a,b],\quad a<b,\quad a,b\enskip \mbox{real numbers}.\eqno{(20)}$$\\

\subsection{The set of all simultaneous \m s}
\noindent
In this subsection we prove the following {\it claim}. Let two compatible observables \$O_A\enskip\Big(=\sum_mo_mE_A^m\Big)\$ and \$O_A'\enskip\Big(=\sum_no_nE_A^n\Big)\$ (cf definition in relation (18)) be given, and let us understand the concept of "refinement" in the improper sense (cf last passage in section 2). Them
an observable \$\bar O_A\enskip\Big(=\sum_lo_lG_A^l\Big)\$ is their {\it common refinement  \IF }  it is a refinement  of the observable \$O_A^M\enskip\Big(=\sum_ko_kE_A^k\Big)\$ defined by relations (19a) and (19b). The latter observable is thus the {\it maximal common refinement} of the given two compatible observables \$O_A\$ and \$O_A'\$.\\

Since any refinement of a refinement is a refinement, also any refinement of a common refinement is a common refinement. Thus {\it sufficiency } easily follows.

To prove {\it necessity}, we assume that an observable \$\bar O_A\enskip\Big(=\sum_lo_lG_A^l\Big)\$ is a common refinement of the given two observables \$O_A\$ and \$O_A'\$. This implies that there are two surjections (onto maps)
$$ \bar f':\enskip\{\forall l\}\enskip\rightarrow\enskip\{\forall m\},\quad
\bar f'':\enskip\{\forall l\}\enskip\rightarrow\enskip\{\forall n\}
 \enskip\mbox{such that}$$  $$
 \forall l:\enskip
 G_A^l\leq E_A^{m\equiv\bar f'(l)},
 \quad
 G_A^l\leq E_A^{n\equiv\bar f''(l)}.\eqno{(21)}$$

Note that \$\forall l:\enskip E_A^{m\equiv\bar f'(l)}
E_A^{n\equiv\bar f''(l)}\not= 0\$ because
\$G_A^l\$ is a non-zero common lower bound of the two factors. Hence we can define an injection (into map) of the index set \$\{\forall l\}\$ into the index set \$\{\forall k\}\$:
$$f\equiv \bar f',\bar f'':
 \forall l:\enskip k(l)
\equiv f(l)=\equiv
[m\equiv\bar f'(l)],[n\equiv\bar f''(l)]$$  $$
\Rightarrow\enskip G_A^l\leq E_A^{k(l)}.\eqno{(22)}$$\\

In section 2 we have seen that over\M is based on measuring a refinement. Relation (22) would prove \$\bar O_A\$ to be a refinement of
\$O_A^M\$ if ut were a surjection of \$\{\forall l\}\$ onto \$\{\forall k\}\$.

In Appendix A it is shown that \$\sum_lG_A^l\leq\sum _kE_A^k\$ (cf relation (A.5)). Since the observable
\$\bar O_A\$ has its completeness relation \$\sum_lG_A^l=I_A\$, we have \$I_A\leq\sum _kE_A^k\$. Since \$I_A\$ is an upper bound of all projectors, we have \$I_A\leq\sum_kE_A^k\leq I_A\$ implying \$\sum_kE_A^k=I_A\$. Hence,  after all, we are dealing with a surjection and the necessity of the claim \$\bar O_A\$ being a refinement of \$O_A^M\$ is proved.\hfill $\Box$\\

Incidentally, the products \$E_A^mE_A^n\$ outside the the image \$\bar f(\{\forall l\})\$ in \$\{\forall m,n\}\$ must be all zero on account of the orthogonality of the eigen-projectors.\\

Every complete observable \$O^C_A\$ (cf (5a-c)) that is a refinement of the maximal common refinement \$O^M_A\$ given by  (19a) and (19b) for two given compoatible observables is a {\it local minimum} in the set of all common refonements. By definiyion this means that \$O^C_A\$ has no refinedment. This is in contrast with \$O^M_A\$, which is a global maximum.\\

\subsection{Corollaries}
\noindent

{\it COROLLARY 1} Let \$\{O_A^q=\sum_{n_q}o_{n_q}E_A^{n_q}:q=1,2,\dots ,Q\}\$ be an arbitrary set of Q (a natural number) pairwise compatible discrete observables in their unique spectral forms. The maximal common refinement \$O_A^M\$ is defined in its unique spectral form as follows.
$$O_A^M\equiv\sum_{n_1}\sum_{n_2}\dots \sum_{n_Q}o_{n_1\dots n_Q}
\prod_{q\in\cQ}E_A^{n_q},\eqno{(23)}$$
where it is understood that all terms in which the projectors multiply into zero are omitted and all eigenvalues are arbitrary but distinct.

Simultaneous \M of all observables from the set is performed \IF \$O_A^M\$ or any of its refinements is measured.\\

{\it PROOF} For \$Q=2\$ the claim has benn proved in the preceding two subsections. Let us assume that its is valid for R observables, where R is a natural number. Then we know, again from the preceding two subsections, that for \$R+1\$ obserbales the claim of Corollapr 1 is valid. Hense, by total induction we conclude that the claim is valid for any natural number Q.\hfill $\Box$\\

{\it COROLLARY 2} Let \$O_A\enskip\Big(=\sum_ko_kE_A^k\Big)\$ be any discrete obserble given in its unique spectral form.
Further, let
$$\{\forall k\}=\sum_l\cC_l \eqno{(24a)}$$
be any breaking up the index set into classes, i. e., writing it as the union of non-intersecting subsets \$\cC_l\$. Then, defining
$$\forall l:\quad E_A^l\equiv\sum_{k\in\cC_l}E_A^k
\eqno{(24b)}$$ any observable
$$\bar O_A\equiv\sum_lo_lE_A^l \eqno{(24c)}$$
with arbitrary but distinct eigenvalues is a coarsening of \$O_A\$ and any \M of \$O_A\$  is, at the same time, also a \m , or rather an over\M of the latter coarsened onservable.\\

No careful reader of section 2 will need proof of Corollary 2.\\

\section{Summing Up}
\noindent
The investigation in this article began with von Neumann's treatment of the \M of any discrete observable via a suitably chosen complete one (cf relations (5a-c)). It was pointed out that the latter observable is a refinement, and its \M is over\M of the initially given observable.

Then a general and detailed theory of over\M was presented in the hope that it will find applications.

Next, the study turned to simultaneous \m , as to an important application of the concept of a refinement of an observable and of over\M as a procedure. It turned out that simultaneous \M is the same thing as common over\m . To illustrate the power of over\M theory, some fine points of simultaneous \m , especially finding the set of all simultaneous \m s for a given pair of observables, have been worked out.\\

{\bf \Large Appendix A: Some helpful projector relations}\\
\noindent
We assume that it is known that the set of all projectors in an at most countably-infinite dimensional complex Hilbert space (state space of a
\Q system) is a partially ordered set with the \q -logical implication \$E\leq F\enskip\Big(\equiv EF=E\Big)\$.
Besides it is a complete lattice, i. e., each non-empty subset has both a greatest lower bound (glb) and a least upper bound (lub). We now prove algebraically a few (more or less well known) claims  that we make use of in subsection 3B.

If the reader knows that there exists a natural isomorphism between the partially ordered set of all projectors and that of all subspaces of the state space, then he may find it easier to supply the proofs in terms of subspaces. (This isomorphism maps a projector
\$E\$ into its range \$\cR(E)\$. The inverse of this isomorphism takes any subspace \cS into the projector  that makes \cS its range.)\\

{\it PROOF} of the claim
$$ EF=FE\quad\Rightarrow\quad EF=glb(E,F).\eqno{(A.1)}$$
Let \$G\$ be any common lower bound of \$E\$ and \$F\$: \$GE=GF=G.\$ Then \$G(EF)=GF=G\$. Thus, \$G\$ is a lower bound a;so of \$EF\$ as claimed.\\

{\it PROOF} of the claim that if
\$\{E_l:l=1,2,\dots ,L\}\$, where \$L\$ may even be the power of a countably infinite set, is a set of pairwise orthogonal projectors, then
$$S\equiv\sum_{l=1}^LE_l=lub\{E_l:\forall l\}.\eqno{(A.2)}$$

The projector S is a common upper bound of the projectors in the given set  because
$$\forall l: E_lS=\sum_{l'=1}^LE_lE_{l'}=E_l.$$
Let \$F\$  be any common upper bound for all projectors
\$E_l\$. Then it is also an upper bound of S because
$$SF=(\sum_{l=1}^LE_l)F=\sum_{l=1}^LE_l=S.$$\\

{\it PROOF} of the claim that if two projectors
\$E,F\$ are orthogonal, and a third projector \$G\$ implies (\q -logically) one of them, then also \$G\$ is orthogonal to the other projector:
$$EF=0,\quad G\leq E, \quad \Rightarrow\quad GF=0.\eqno{(A.3)}$$
This is so because
$$GF=(GE)F=G(EF)=0.$$\\

{\it PROOF} of the claim that if one has two finite or infinite sums of pairwise orthogonal projectors, \$(\sum_lG_l,\sum_kE_k)\$ such that a map \$f\$ is given that takes the index set \$\{l\}\$ into the index set \$\{k\}\$ so that
$$\forall l:\quad G_l\leq E_{k=f(l)},\eqno{(A.4)}$$
then the former sum is a lower bound of the latter
$$\sum_lG_l\leq\sum_kE_k.\eqno{(A.5)}$$

To begin the proof, we single out the subset \$\{\bar k=f(l):\forall l\}\enskip\Big(\subseteq\{\forall k\}\Big)\$ that is the image of \$\{\forall l\}\$ regarding the map \$f\$. Next we break up \$\{\forall l\}\$ into classes \$\cC_{\bar k}\equiv f^{-1}(\bar k)\$:
\$\{\forall l\}=\sum_{\bar k}\cC_{\bar k}\$. Then. we claim that
$$\forall \bar k:\quad \cC_{\bar k}E_{\bar k'}= \delta_{\bar k,\bar k'}\cC_{\bar k}.\eqno{(A.6)}
$$
To prove the step (A.6), one has
 \$\cC_{\bar k}E_{\bar k}=\sum_{l\in f^{-1}(\bar k)}G_lE_{\bar k}=
\cC_{\bar k}\$ due to (A.4). As to the claimed (logical) implication, \$(\bar k\not=\bar k')\enskip\Rightarrow\enskip \cC_{\bar k}E_{
\bar k'}=0\$, it follows from (A.3) because \$E_{\bar k}E_{\bar k'}=0\$, and \$\cC_{\bar k}\leq E_{\bar k'}\$.

Relation (A.6) implies
$$\sum_lG_l=\sum_{\bar k}\cC_{\bar k}\leq \sum_{\bar k}E_{\bar k} \eqno{(A.7)}$$
because \$(\sum_lG_l)(\sum_{\bar k}E_{\bar k})=
\sum_{\bar k,\bar k'}\cC_{\bar k}E_{\bar k}=
\sum_{\bar k}\cC_{\bar k}=\sum_lG_l\$.

Next, the orthogonality of the projectors \$E_k\$ implies
$$\sum_{\bar k}E_{\bar k}\leq \sum_kE_k=
\sum_{\bar k}E_{\bar k}+\dots .\eqno{(A.8)}$$
Hence, the transitivity of \q -logical implication
supplies the final proof of (A.5) \$\sum_lG_l\leq\sum_kE_k\$.\\

{\bf \Large Appendix B: On compatibility of bounded observables}\\
\noindent
The main claim of Appendix B is a consequence  of the following {\it more general claim}:

{\it CLAIM 1.} Let \$O=\sum_ko_kE_k\$ be a bounded discrete observable in its unique spectral form and let \$\bar O\$ be a bounded linear operator. Then the following three relations are {\it equivalent}:
$$(1)\enskip [O,\bar O]=0\quad\Leftrightarrow\enskip  (2)\enskip \bar O=\sum_kE_k\bar OE_k$$  $$\Leftrightarrow\enskip  (3)\enskip  \forall k:\enskip [E_k,\bar O]=0.\eqno{(B.1)}$$\\

{\it PROOF} of the claimed (logical) implications in (B.1) will be given in an in-circle way as follows: (1) $\Rightarrow$ (2) $\Rightarrow$ (3) $\Rightarrow$ (1).

\$(1)\Rightarrow (2)\$: One can write (1) in (B.1), on account of
\$I=\sum_kE_k\$,  as follows, and, multiplying out the factors in the commutator, one obtains
$$[\sum_ko_kE_k,\sum_k\sum_{k'}E_k\bar O E_{k'}]=0\enskip\Rightarrow$$  $$ \sum_k\sum_{k'}o_kE_k\bar OE_{k'}\enskip -\enskip
\enskip \sum_k\sum_{k'}o_{k'}E_k\bar OE_{k'}=0$$  $$\Rightarrow\enskip \sum_{k\not= k'}(o_k-o_{k'}) E_k\bar OE_{k'}=0.\eqno{(B.2)}$$

Taking fixed \$k\$ and \$k'\$, we multiply the last relation by \$E_k\$ from the left and by \$ E_{k'}\$ from the right to obtain \$(o_k-o_{k'}) E_k \bar O E_{k'}=0\$, and finally \$\forall (k\not= k'):\enskip E_k\bar OE_{k'}=0\$. Thus, (2) follows from (1).\\

\$(2)\Rightarrow (3)\$: Multiplying \$\bar O=\sum_k E_k\bar OE_k\$ from the left or alternatively from the right by the same arbitrary fixed \$E_k\$, one obtains the same term \$E_k\bar OE_k\$. Hence (3) is a consequence of (2) in (B.1).\\

\$(3)\Rightarrow (1)\$: The third relation implies
$$[O,\bar O]=\sum_ko_k[E_k,\bar O]=0.
 \eqno{(B.3)}$$ This ends the proof.\\
 
Claim 1 implies the claim that we actually want to prove in this appendix.\\
{\it CLAIM 2.} Let \$O=\sum_mo_m E_m\$ and \$O'=\sum_no_nE_n\$
be two discrete Hermitian operators given in their unique spectral forms.
Then the two operators commute, \$[O,O']=0\$,
\IF each eigen-projector of the former commutes with each eigen-projector of the latter \$\forall m,n:\enskip [E_m,E_n]=0\$.\\

{\it PROOF. Sufficiency.} Assuming \$\forall\enskip m,n:\enskip [E_m,E_n]=0\$, one obtains \$[O,O']=0\$ as seen by substituting the unique spectral forms for both operators and utilizing the bilinearity of the commutator.

{\it Necessity.} According to the above proposition, (1)in (B.1) implies (3) in (B.1). Hence, \$[O,O']=0\$ implies \$\forall m:
\enskip [\hat E_m,O']=0\$. A repeated application of the mentioned claim in the Proposition  can now be written as
$$\forall m:\enskip [O',E_m]=0\enskip\Rightarrow\enskip
\forall n:\enskip [E_n,E_m]=0$$.\hfill $\Box$\\

\end{document}